\documentclass[12pt,preprint]{aastex}

\usepackage{psfig}  

\def\e{{\rm E}}
\def\rel{{\rm rel}}
\def\thr{{\rm thr}}
\def\Sb{\Sigma_{\bullet}}

\newcommand{\etal}{{et al.}\ }

\newcommand{\kpc} {$\, {\rm kpc}$}
\newcommand{\kms} {$\, {\rm km/s}$}

\newcommand{\msun}{M$_{\odot}$}

\begin{document}

\def\newpage{\vfill\eject}
\def\vs{\vskip 0.2truein}
\def\pp{\parshape 2 0.0truecm 16.25truecm 2truecm 14.25truecm}
\def\fun#1#2{\lower3.6pt\vbox{\baselineskip0pt\lineskip.9pt
  \ialign{$\mathsurround=0pt#1\hfil##\hfil$\crcr#2\crcr\sim\crcr}}}
\def\core{{\rm core}}
\def\min{{\rm min}}
\def\max{{\rm max}}
\def\kpc{{\rm kpc}}
\def\esc{{\rm esc}}
\def\crit{{\rm crit}}
\def\pc{{\rm pc}}
\def\kms{{\rm km}\,{\rm s}^{-1}}
\def\cbh{{\rm cbh}}
\def\bh{{\rm bh}}
\def\df{{\rm df}}
\def\bulge{{\rm bulge}}

\shortauthors{Chanam\'e et al.\ }
\shorttitle{Microlensing by the Sgr A* Cluster of Black Holes}
\title{Microlensing by Stellar Black Holes Around Sgr A*}
\author{Julio Chanam\'e\,$^1$, Andrew Gould\,$^{1,2}$, \& Jordi Miralda-Escud\'e\,$^{1,3}$} 
\affil{{}$^1$ Department of Astronomy, The Ohio State University,
Columbus, OH 43210, USA}
\affil{{}$^2$ Laboratoire de Physique Corpusculaire et Cosmologie, 
Coll\`ege de France, F-75231, Paris, France}
\affil{{}$^3$ Alfred P. Sloan Fellow}
\email{jchaname@astronomy.ohio-state.edu, gould@astronomy.ohio-state.edu,
jordi@astronomy.ohio-state.edu}

\begin{abstract}

We show that at any given time, the Galactocentric black hole Sgr A*
is expected to be microlensing $N_{\rm lens}\sim 1.7$ bulge stars if
the threshold of detectability of the {\it fainter} image is
$K_\thr=21$, and $N_{\rm lens}\sim 8$ sources if $K_\thr=23$. The
lensed images then provide a unique way to detect stellar-mass black
holes predicted to cluster around Sgr A*. If a black hole passes close
to a microlensed image, it will give rise to a short (weeks long)
microlensing event.  We show that the mass and projected velocity of
the stellar-mass black hole can both be measured unambiguously from
such an event, provided that either a caustic crossing is observed or
the astrometric displacement is measured.  For $K_\thr=23$ and
moderate magnifications by Sgr A*, the microlensing event rate from a
cluster of 20000 black holes within a radius of 0.7 pc is only
$0.06\,{\rm yr}^{-1}$; however, if highly magnified images of a star
were found, the rate of events by the stellar black holes would be
much higher.  In addition, the $N_{\rm lens}$ sources lensed by Sgr A*
provide a unique probe of extinction {\it behind} the Galactic center
along $2N_{\rm lens}$ lines of sight.

\end{abstract}

\keywords{black hole physics -- 
Galaxy: center -- Galaxy: kinematics and dynamics -- gravitational lensing}

\setcounter{footnote}{0}
\renewcommand{\thefootnote}{\arabic{footnote}}

\section{Introduction}

Any black holes formed from the final core collapse of massive stars
within the central 5 pc of the Milky Way during the last
10 Gyr should have migrated to the Galactic center (GC) owing to dynamical
friction off ordinary stars (Morris 1993). 
Miralda-Escud\'e \& Gould (2000, hereafter MG) found that the majority
of these black holes survive capture by the massive black hole
Sgr A* and should still be present in the GC today.
Low-mass stars older than $\sim 1\,$Gyr should have been expelled from
this region by the relaxation process. If stars with an initial mass
greater than $\sim 30\,M_\odot$ generally produce black holes,
then this black-hole cluster should contain about 20,000 members today.

  One of the methods to test for the existence of the black-hole
cluster is microlensing of a background source, probably
a bulge star.  This method requires first the discovery of the two 
images of a bulge star lensed by Sgr A*, and then the monitoring of 
these images to detect the short-duration (weeks long) microlensing events  
caused by the passage of one of the cluster black holes near an image.

  The observational challenge here is to identify the two images of a
faint $(K\sim 21)$ lensed star in a field crowded with $\sim 400\,\rm
arcsec^{-2}$ stars of similar magnitude and brighter (see \S\ 2.1).
Hence, a resolution $\ll 0.\hskip-2pt'' 1$ is required.  Such deep,
high-resolution observations are beyond current capabilities, but
may be achievable with improvements in adaptive optics allowing long
exposures, and with NGST. It will then be possible to distinguish the
pair of resolved images from the more numerous field stars by means of
their positions, flux ratio, colors and proper motions,
which are all related.  We show in \S\ 2, that there are
likely to be a few such detectable image pairs at any given time.
However, as we show in \S\ 3, the rate of events caused by cluster
black holes is only about 1 per 100 years per image pair, so that
still deeper observations will probably be required to detect
stellar black holes.  In \S\ 4, we show that the
masses of these black holes could be measured either by photometric
monitoring of a caustic crossing, or from their astrometric effects.
Finally, in \S\ 5, we discuss implications of our results.

Alexander \& Sternberg (1999) calculated lensing rates for Sgr A*
considered as an isolated body, and obtained results that are broadly
consistent with those presented here.  The problem of microlensing of
background disk sources by stars orbiting near Sgr A* (similar to the
problem we treat here) was investigated by Alexander \& Loeb (2001).


\section{Expected number of stars lensed by Sgr A*}

The expected number of sources being lensed by Sgr A* at any given
time can be estimated empirically given three observational inputs:
1) the $K$-band luminosity function (hereafter, LF) of the sources,
2) the volume
density of $K$-band light as a function of distance behind the GC, and
3) the extinction as a function of distance behind the GC.  The first
two of these are well determined from observations, as we summarize
below.  The last is not measured.  For purposes of this paper we will
assume that there is no significant additional extinction in the $\sim
1\,$kpc lying behind the GC, above and beyond the $A_K=3$ magnitudes
of extinction known to lie in front of the GC (e.g., Blum, Sellgren,
\& DePoy 1996).

\subsection{Distribution of Stars behind the Galactic Center}

We derive the (dereddened) bulge $K_0$ LF from the $J$-band LF
measured by Zoccali et al.\ (2000, hereafter Z00; see their Table 1)
using the {\it Hubble Space Telescope} NICMOS $J$ and $H$ images of a
field lying at $(l,b)=(0^\circ ,-6^\circ )$ (with a field size
$\Omega_{\rm Zocc} = 22.\hskip-2pt''5 \times 22.\hskip-2pt''5$). To
obtain the complete $J$-band LF, Z00 combined their NICMOS
observations, which yield the LF from $J$=16.5 to $J$=24.5, with the
bright LF from the Baade's Window data of Tiede, Frogel, \& Terndrup
(1995) which extends up to the tip of the red giant branch.  We
convert the NICMOS part of this LF to $H_0$ using the reported
extinction and observed $(J,J-H)$ color-magnitude diagram in Z00, and
then convert to $K_0$ using the $(K,H-K)$ relation observed for nearby
main-sequence stars (e.g.,\ Henry \& McCarthy 1993).  For the bright
extension of the LF, we convert to $K_0$ using the $(K,J-K)_0$
color-magnitude diagrams reported by Tiede et al. (1995, Figs. 6 and
7).  The resulting LF, expressed as the number of stars per square arc
minute per K apparent magnitude (i.e., including our adopted
extinction to the GC), is shown in Figure 1.  It reaches $K\sim 26.5$,
and includes bulge main sequence stars with masses $M\ga
0.15\,M_\odot$.  Note that the LF for main-sequence stars is nearly
constant throughout the entire bulge (Narayanan, Gould, \& DePoy
1996), while luminous stars are more abundant in the GC; hence, this
extension of the LF will underestimate the number of lensed luminous
stars close to the GC.

\vspace{0.3cm}
\centerline {EDITOR: PLEASE PLACE FIG. 1 HERE}  
\vspace{0.3cm}  

     To convert this to a local {\it volume density} $K$ LF,
$(d N/d K d V)(D_{s})$, at a distance $D_{s}$ along the line of sight of 
Sgr A*, we first write
\begin{equation}
{d N\over d K d V}(D_{s}) = 
{\rho(D_{s})\over\Sigma_{\rm Zocc}}\,
{(d N_{\rm Zocc}/d K_0)|_{K_0 = K - \Delta K}
\over R_0^2\Omega_{\rm Zocc}},
\label{eqn:conv1}
\end{equation}
where $\rho(D_{s})$ is the local bulge mass density on the GC line of
sight at a distance $D_s$ from us, $\Sigma_{\rm Zocc}$ is the bulge
column density toward the Z00 field, $dN_{\rm Zocc}/dK_0$ is the
number of stars per $K_0$ magnitude in the Z00 field, $R_0=8\,$kpc is
the Galactocentric distance, $\Omega_{\rm Zocc}$ is the solid angle of
the Z00 field, $\Delta K=A_K(D_{s}) + 2.5\log(D_{s}/R_0)$, and where
we assume $A_K(D_{s})=3$.  In practice, we set $\Delta K = A_K = 3$,
thus neglecting the dimming due to larger distances, since this
contribution is smaller than the uncertainty introduced by assuming a
constant $A_K(D_{s})$.

  The first factor on the right-hand side of equation (\ref{eqn:conv1})
depends only on the spatial distribution of the mass in the bulge, and
not its normalization. To evaluate it, we use a combination of the Kent
(1992) model, which is based on K band images, and the DIRBE data from
Dwek et al.\ (1995). The reason this combination is required is that
the Kent model is axisymmetric, and so does not include the effects of
the bar which are important at large radius, while the DIRBE map cannot
be applied at small radius because of its limited resolution. We rewrite this
term as
\begin{equation}
{\rho(D_{s})\over\Sigma_{\rm Zocc}} = 
{\rho(D_{s})\over\Sigma_{(0,-2)}}\,
{\Sigma_{(0,-2)}\over\Sigma_{\rm Zocc}},
\label{eqn:conv2}
\end{equation}
where $\Sigma_{(0,-2)}$ is the column density in the direction
$(l,b)=(0,-2^\circ )$. We determine the first ratio from equation (3)
in Kent (1992), except that we impose a constant density core within
0.7 pc due to relaxation effects around Sgr A* (MG).  We determine the
second ratio from the DIRBE map (Fig.\ 1a in Dwek et al.\ 1995),
obtaining a value $\Sigma_{(0,-2)}/\Sigma_{\rm Zocc}\simeq 13$.  Note
that this procedure is relatively insensitive to possible disk
contamination of the Z00 field since to leading order this
contamination affects both the star counts and the DIRBE map equally.
We have also used this same procedure to predict the surface density
of stars near Sgr A*, which is found to be 440 arcsec$^{-2}$ stars
with $K<21$ mag, and about 1000 arcsec$^{-2}$ with $K<23$ mag.

\subsection{Number of Lensed Sources}

	The two images of a microlensed source will have
magnifications given by
\begin{equation}
A_\pm(u) = {u_\pm^2\over u_+^2-u_-^2},\quad
u_\pm = {u\pm\sqrt{u^2+4}\over 2},\quad
u\equiv {\theta_\rel\over\theta_\e},
\label{eqn:apm}
\end{equation}
where $\theta_\rel$ is the angular separation of the lens and source,
$\theta_{I\pm}=u_\pm\theta_\e$ are the positions of the two images,
$\theta_\e$ is the Einstein radius,
\begin{equation}
 \theta_\e(D_s) = \sqrt {\frac{4GM_\ast}{c^{2}} \frac{D_{ls}}{D_{l}D_{s}}} =
0.\hskip-2pt'' 55 \,
\biggl({10 D_{ls} \over D_s }\biggr)^{1/2},
\label{eqn:thetaeofd}
\end{equation}
$M_\ast=3\times 10^6\,M_\odot$ 
is the mass of Sgr A*, $D_{l}=R_0$, $D_{s}$ is the distance
to the source, and $D_{ls}=D_s-D_l$.
The angular separation of the two lensed images of a background star at
$D_{ls}\sim $ 1 kpc is $\sim 2\theta_\e\sim 1''$, clearly large enough to 
resolve them. On the other hand, the duration of the ``microlensing event'' 
is very long, given the typical proper motion of bulge stars of a few mas 
per 
year.  This implies
that the usual method of microlensing detection from 
the magnification lightcurve is impractical.  Thus, to detect a lensed
source, it is essential to identify {\it a pair of images}, which can be
done from their relative magnifications 
and proper motions (both of which are unambiguously predicted from their
positions relative to Sgr A*), as well as their common dereddened color.

	The expected number of observable lenses $N_{\rm lens}$ is simply the 
number for which the fainter image is brighter than the threshold of 
detectability, $K_\thr$,
\begin{equation}
N_{\rm lens} = \int_{R_0+0.7\,\rm pc}^{D_{s,\rm max}} d D_s\,D_s^2
\int_0^\infty d\theta\,2\pi\theta
{d N\over d V}\biggl[D_s,K_\thr 
           + 2.5\log A_-\biggl({\theta\over\theta_\e(D_s)}\biggr)\biggr],
\label{eqn:nlens}
\end{equation}
where $dN/dV(D_s,K')= \int^{K'} dK\,(dN/dVdK)(D_s,K)$ is
the cumulative LF. We use $D_{s,\rm max}=9$ kpc, effectively assuming that
additional obscuration becomes important farther than 1 kpc behind the GC.
We find for $K_{\rm thr}=(21,22,23)$, that
$N_{\rm lens}=(1.7,3.8,8.0)$, which can be summarized,
\begin{equation}
N_{\rm lens} \simeq 8\times10^{0.33(K_{\rm thr} - 23)},
\qquad (21<K_{\rm thr}<23).
\label{eqn:nlenseval}
\end{equation}
Thus, in order to have a reasonable probability of finding one lensed
star, the inner $1''$ of the GC must be imaged down to at least $K\sim
21$.  Once a lensed image pair is detected, it is mainly the brighter
member that will be useful to search for black-hole cluster members.
Figure 1 shows the distribution of brighter and fainter members (solid
and dashed lines, respectively) for the three cases
$K_\thr=(21,22,23)$.  From 0.7 pc and up to 1 kpc behind the GC the
contributions of equal logarithmic intervals of $D_{ls}$ are 24\%,
31\%, and 45\%.  Therefore, although most lensed sources will be from
the ``outer'' bulge, with $\theta_\e\ga 0.\hskip-2pt'' 1$, a
significant fraction will be close to Sgr A*, with $\theta_\e\sim
30\,$mas ($\sim$ 250 AU), corresponding to $D_{ls}\sim$ 3 pc.  Hence,
detection of these images, at $\theta_\e\sim 30\,$mas, might be
compromised by extra extinction or emission from an accretion
flow/disk around Sgr A*.

	In the above calculations it has been assumed that there is essentially
no extra extinction in the 1 kpc behind the GC.  In fact, most of the
extinction measured towards the GC is thought not to arise in the
immediate vicinity of the Galactic center, but several kpc away, in a
ring of molecular material.  If this holds behind the GC as well, then our
results for $N_{\rm lens}$ are not affected.  On the other hand, if
the situation behind the GC is different than in front of it, there
are two possibilities.  First, if, say, $A_K=1$ magnitude of extinction is 
uniformly
distributed across the first kpc behind, then the closer sources will
be the main contributors to $N_{\rm lens}$, and the numbers quoted in
equation (6) would be reduced by about 30-40\%.  A second possibility
is that the extra extinction is concentrated close to the GC, in which
case all the sources behind are equally affected.  Hence, if
there is 1 mag of extra extinction close to the GC, then the
quoted numbers for $N_{\rm lens}$ would be for $K_{\rm thr}$ =
(22,23,24), instead of (21,22,23).  The detection of lensed sources will
itself allow a quantitative probe of extinction behind the GC, as we
discuss in \S\ 5.

In Figure 2 we show as solid lines the shear distribution
$(dN/d\gamma)$ of the brighter pair member obtained for the three
magnitude limits.  Pairs with high magnifications (shear close to
unity) are less numerous than those with small magnifications.  The
area under any of these curves corresponds to the number of lensed
pairs that are expected to be found at the corresponding detection
threshold.  The solid line in the left hand panel of Figure 3, below,
shows $N(>\gamma)$, the cumulative distribution of $\gamma$ of the
brighter pair member, for the case $K_{thr} = 23$.

\vspace{0.3cm}
\centerline {EDITOR: PLEASE PLACE FIG. 2 HERE}
\vspace{0.3cm}

\section{Rate of Microlensing Events by the Cluster Black Holes}

After two images of a background bulge star lensed by Sgr A* have been
identified, monitoring them could reveal short microlensing events due
to one of the cluster black holes passing near one of the images.  If
the typical cluster black hole has a mass comparable to the masses
inferred from X-ray binary systems, $m_{\rm bh}\simeq 7$ \msun (Bailyn
et al.\ 1998), the mass ratio of the binary lens of Sgr A* and the
cluster black hole is $q= m_{\rm bh}/M_{\ast} \sim 2\times 10^{-6}$.
The lightcurves would be similar to those caused by planets around
ordinary stars (Mao \& Paczy\'nski 1991; Gould \& Loeb 1992; Albrow et
al.\ 1998), although with longer timescales. In this section, we
calculate the rate at which these planet-like events should take
place.

  The rate of microlensing due to cluster black holes depends on their
surface density, their velocity distribution, 
and a {\it linear cross section} giving the size of
the region around an unperturbed image where a cluster black hole will
significantly change its magnification.
We obtain the surface density of black holes, $\Sb(b)$ (where $b$ is the
projected radius), using the model in
MG, where the cluster density varies as
$\rho(r)\propto r^{-7/4}$, the total stellar mass enclosed within
$r_0=1.8$ pc is equal to $M_{\ast}=3\times 10^6$ \msun, and the density
normalization of the black hole cluster is reduced by a factor
$(m_{\rm bh}/\overline{m})^{-1/2} = 0.18$
below that of the cluster of stars, owing to the change in the mean mass
of the objects between the black hole cluster $(m_{\rm bh}\sim 7\,M_\odot)$
at $r \lesssim 0.7$ pc
and the stellar cluster at larger distances
$(\overline{m}\sim 0.23\,M_\odot)$.
At projected radii 
$b \ll 0.7$ pc, the result is 
\begin{equation}
\Sb(b) = 0.18 \,\frac{5}{16 \pi} \,
{M_{\ast}/m_{\rm bh}\over {r_{0}^{5/4}}}
\int_{-\infty}^{\infty} (z^2 + b^2)^{-7/8}\, dz = 
1.6\times 10^{5} \,
\biggl({\theta\over 1''}\biggr)^{-3/4} {\rm pc^{-2}} \, ,
\end{equation}
where $\theta\equiv b/R_0$ is the angular distance from the center. 

  We note here that the reduction factor
$(m_{\rm bh}/\overline{m})^{-1/2} = 0.18$ probably gives us an
underestimate of the surface density of lensing objects, because the
time necessary for the black hole cluster to expand and achieve
equilibrium under relaxation with the low-mass stars is somewhat longer
than the age of the universe (this was not taken into account in MG).
This means that the rate of events we will find may be underestimated,
proportionally to the surface density. This issue will be addressed
in a future paper.

  We assume an isotropic velocity distribution.
For $\rho(r) \propto r^{-7/4}$, the Jeans equation
(see Binney \& Tremaine 1987) yields a one-dimensional velocity dispersion 
$\sigma(r) = \sqrt{4/11} \; v_{c}(r) = 68.5 (r/{\rm pc})^{-1/2} \,\kms$, where
$v_c(r)$ is the circular velocity. The projected
two-dimensional velocity dispersion is $\sigma_p(b)=0.99 \sigma(b) =
68 (b/{\rm pc})^{-1/2} \,\kms$.

We calculate the linear cross section assuming a Chang-Refsdal lens
(Chang \& Refsdal 1979; Schneider, Ehlers \& Falco 1992), of a point
mass in the external shear $\gamma=(\theta_\e/\theta_I)^2$ caused by
Sgr A*, where $\theta_I$ is the angular distance from the image to Sgr
A*. We first consider events on the brighter member of the unperturbed
pair, with unperturbed magnification $A_+$ given by equation (2).  We
will later include the contribution from the fainter images.  Our
criterion for the detectability of the short microlensing event is
that the magnification of this image is increased to at least
$(1+\delta) A_+$. If $\xi$ is the axis joining Sgr A* and the cluster
black hole, and $\eta$ the perpendicular axis, we find that the
contours of constant $\delta$ in the source plane have lengths along
the $(\xi, \eta)$ axes of
\begin{equation}
L_{\xi} = \frac{ 2\,(2\gamma -\lambda +1)}{\sqrt{\lambda -\gamma} }
          \sqrt{q}R_0\theta_\e \, ,
~~~~~
L_{\eta} = \frac{ 2\,(2\gamma +\lambda -1)}{\sqrt{\lambda +\gamma} }
           \sqrt{q}R_0\theta_\e \, ,
\label{lincr}
\end{equation}
\noindent where 
\begin{equation}
\lambda=\left(1-\frac{1}{(1+\delta)A_{+}} \right)^{1/2}.
\end{equation} 
We first consider the case $\lambda=1$, which corresponds to requiring a
caustic crossing (i.e., $\delta = \infty$).

  To compute the rate of events, we notice first that the surface
density of caustics in the source plane is $\Sb/(1-\gamma^2)$, and the
velocity dispersion of these caustics is anisotropic. To render the
calculation more transparent, it is convenient to make a linear
transformation of the source plane to $\xi'=\xi/(1+\gamma)$, and
$\eta'=\eta/(1-\gamma)$
(this transformation would be locally equivalent to the lens plane in
the absence of the perturbing black hole). In this transformed source
plane, the surface density and velocity dispersion of the caustics are
the same as the surface density and velocity dispersion of the black
holes in the lens plane, and the dimensions of contours of constant
$\delta$ are modified to $L_{\xi'} = L_{\xi}/(1+\gamma)$,
$L_{\eta'} = L_{\eta}/(1-\gamma)$.

The rate of events for a given unperturbed image is just the product
$\,\omega = \Sb\, \sigma_p\, \overline{L}'$, where $\overline{L}'$ is
the average length of the $\delta$ contour over all possible
directions of motion of the black hole, and where $\Sb$ and $\sigma_p$
are evaluated at the projected radius of the brighter unperturbed
image, $b=\gamma^{-1/2} R_0\theta_\e$. In the case $\delta=\infty$,
$\overline L'$ can be evaluated by noticing that, because the shape of
the caustic is always concave between cusps, the cross section for any
angle of motion is determined solely by the position of the four
cusps.  If $\alpha$ is the angle of the motion relative to the $\xi'$
axis, the cross section is $L_{\xi'}\cos\alpha$ for $\alpha <
\alpha_c$, and $L_{\eta'}\sin\alpha$ for $\alpha > \alpha_c$, where
$\tan\alpha_c = L_{\xi'}/L_{\eta'}$. Hence,
\begin{equation}
\overline{L}' = \frac{2}{\pi}\, \left(\int_{0}^{\alpha_c}\, 
L_{\xi'}\, \cos\alpha \,\, d\alpha \; + \int_{\alpha_c}^{\pi/2}\,
L_{\eta'}\, \sin\alpha \,\, d\alpha \right) =  \frac{2}{\pi}\,\sqrt{L_{\xi'}^2
+L_{\eta'}^2} = \frac{8\,\sqrt{2}\,\gamma}{\pi(1-\gamma^2)}\sqrt{q}R_0\theta_\e
\,.
\end{equation}
This yields an event rate
\begin{equation}
\omega = 1.2 \times 10^{-2}\, \frac{\gamma^{13/8}}{1-\gamma^2}\,
\biggl({\theta_\e\over 1''}\biggr)^{-1/4}\, {\rm yr}^{-1}\, .
\label{rateim}
\end{equation}
For a bulge star at $D_{ls}$=500 pc (i.e., $\theta_\e \simeq
0.\hskip-2pt ''4$), a shear of $\gamma$=0.6 (i.e., impact parameter
$u$=0.52, $A_{+}$=1.56), this implies approximately 0.01 microlensing
events per year on the brighter image. If a very highly magnified pair
of images is found, the rate of events is enhanced. For example, for
$\gamma=0.97$, when the brighter image has an unperturbed
magnification of 17, the rate is enhanced to $\sim 0.25$ events per
year (an event like this would last for 2 months, while the total Sgr
A* event has an Einstein radius crossing time of 100 years).

  We can now compute the rate of events as a function of $\gamma$,
by multiplying the mean number of images lensed by Sgr A* (which is
shown as the solid lines in Figure 2) times the event rate on a given
image from equation (\ref{rateim}). The result is shown as the dashed
line in Figure 2. The cumulative event rate is shown in the left hand
panel of Figure 3 as the dashed curve. A large fraction of the mean
rate of events occur on high magnification images. The mean rate of
events (integrated over $\gamma$) is (0.012, 0.021, 0.034) yr$^{-1}$,
for $K_{thr}$=(21, 22, 23), respectively. However, the actual expected
rate depends strongly on whether images of high $\gamma$ (lensed by
Sgr A*) are found. For example, the expected number of lensed
images for $K_{thr}=23$ is 8, of which an average of 1 has
$\gamma > 0.7$, and two thirds of the stellar black hole events are
contributed by images with $\gamma > 0.7$.

\vspace{0.3cm}
\centerline {EDITOR: PLEASE PLACE FIG. 3 HERE}
\vspace{0.3cm}

  So far, we have calculated only the rate of {\it caustic
crossing} microlensing events.  In fact, many events can be detected
that do not cross the caustic, although as we discuss in \S\ 4, the
interpretation of these may be more difficult.  The precise
requirement for detection will depend on the precision with which the
flux from these faint images could be monitored.  For illustration,
the dotted line in Figure 4 shows the rate of events $\Gamma(\gamma)$
for $\delta=1$ (i.e., when the unperturbed image is required to be
magnified by a factor of at least 2 relative to its magnification
due to Sgr A* only), compared to $\delta=\infty$, for $K_{thr}=23$.
The total rates are almost doubled, to (0.022, 0.038, 0.062) yr$^{-1}$,
for $K_{thr}$=(21, 22, 23).

\vspace{0.3cm}
\centerline {EDITOR: PLEASE PLACE FIG. 4 HERE}
\vspace{0.3cm}

  The characteristic timescale of the events is 
$\Delta t \sim (0.5\,L_{\xi'}L_{\eta'})/(2 \,\overline{L}' \sigma_p)$, 
so for caustic crossing we obtain
\begin{equation}
\Delta t \,\sim \, 68\, \frac{\gamma^{3/4}}{(1-\gamma^{2})^{1/2}}\,
\biggl({\theta_\e\over 1''}\biggr)^{3/2}
\biggl({m_{\rm bh}\over 7\,M_\odot}\biggr)^{1/2} ~ {\rm days}\, .
\end{equation}
With $\gamma=0.6$ and $\theta_\e \simeq 0.\hskip-2pt ''4$, this gives
an event duration of $\sim$ 2 weeks.  High magnification events last
longer, taking $\sim$ 2 months to complete a $\gamma$ = 0.97 event.

The Chang-Refsdal approximation used here is potentially subject to
severe problems when approaching the limit of very high magnifications
($\gamma\rightarrow$ 1).  This is because there is a point where the
caustics become so large that they begin to overlap, so that we reach
the ``optically thick'' lensing regime in which several cluster black
holes are significantly affecting the magnification of the unperturbed
image at the same time.  Also the caustics weaken in this regime, so
that when the finite size of the source is taken into account, the
caustic crossing may be difficult to detect.  We estimate this to
happen at $\gamma\simeq$ 0.98 (at which \,\,$\omega\,\Delta t\,
\simeq$ 0.1).  The vertical dotted line in the left hand panel of Figure 3
marks this point, beyond which the contribution to the rate is only
$\sim$ 8\% of the total.  While images of background sources highly
magnified by Sgr A* are easier to identify, the interpretation of
microlensing events on such images gets complicated because of the
merging of separated caustics.  Hence it is desirable to wait for the
identification of a lensed image pair and then, if the corresponding
shear is in this regime, work the actual case through more carefully.

In our calculation of the rate of events we have so far considered
only the brighter image of each pair of stars lensed by Sgr A*.
Figure 2 shows that high magnification pairs, though rare, dominate
the rate of microlensing events.  One can see from Figure 3 that
approximately 50\% of the total rate comes from images with
$\gamma>0.8$, for which the ratio of their magnifications to that of
their corresponding fainter images is ${\cal R} =1/\gamma^{2} <$ 1.56.
Thus, the events on the fainter image should be of comparable
importance.  The precise calculation of the cross section for the
fainter images is more difficult because the caustics have a different
morphology for $\gamma > 1$ (two triangles) than for $\gamma < 1$ (one
quadrilateral).  However, for $\gamma\rightarrow$ 1 (which is of
relevance here), the combined cross section of the two triangles is
very well approximated by that of a single quadrilateral with total
extent equal to that of the two triangles (see Schneider, Ehlers, and
Falco, 1992, p.\ 259).  We therefore calculate the rate using this
approximation and find, for events on the fainter images,
(0.008,\,\,0.015,\,\,0.024) yr$^{-1}$, for $K_{thr}$=(21,\,22,\,23)
respectively.  We show the result for $K_{thr}$=23 in the right panel
of Figure 3.

\section{Measuring the Black Hole Masses}


  The stellar black holes have masses much smaller than that of Sgr A*, 
and so are formally identical to planetary lenses. In the case of
planetary systems, the orbital velocity of the planet is usually
negligible compared to the relative velocity of the lens and the source,
and so the mass ratio can be determined from the lightcurve alone by
measuring the ratio of timescales of the planetary perturbation and the
entire event (Gould \& Loeb 1992). However, in the present case the
orbiting black holes have speeds that are the same order or larger
than the sources being lensed.

  It is nevertheless often possible to recover the black-hole mass.
For caustic-crossing events, this can be done just from photometry.
In principle, the black-hole mass can also be recovered astrometrically,
both for caustic-crossing and non-caustic crossing events.  However,
for non-caustic crossers, this determination will prove difficult in
practice.

  First, note that
the normalized lens-source separation, $u$, can be determined either
from the flux ratio $\cal R$ of the two images, 
$u={\cal R}^{1/4}-{\cal R}^{-1/4}$,
or from the measured fractional offset 
$\Delta\equiv(\theta_{I+}+\theta_{I-})/(\theta_{I+}-\theta_{I-})$ 
of Sgr A* from the midpoint of the two images, 
$u=2(1/\Delta -1)^{-1/2}$.  One can then determine the
Einstein radius, 
$\theta_\e = (\theta_{I+}-\theta_{I-})/({\cal R}^{1/4}+{\cal R}^{-1/4})$, or
$\theta_\e = \sqrt{\Delta^{-1}-\Delta}(\theta_{I+}-\theta_{I-})/2$. From 
the known mass of Sgr A*, this then determines the distance to the source.
Next, the shear $\gamma = {\cal R}^{-1/2}$ or 
$\gamma = (1-\Delta)/(1+\Delta)$
determines the geometry of the 
Chang-Refsdal lens, up to an overall scale factor $q^{1/2}$ (proportional
to the size of the caustic).  The source speed is known from the proper
motion of the major image across the sky, thus, the only two unknowns are 
the velocity
of the lens (which determines how fast the caustic moves over the source),
and the size of the caustic, related to the mass of the lens. Measuring
the mass of the lens is the crucial step required to prove that a
microlensing event is caused by a stellar black hole.

If the source passes through the caustic, then its trajectory can
be determined from the photometric light curve.  This will be much
easier for black-hole caustics than for caustics that have been
analyzed in binary-lens events to date (e.g., Afonso et al.\ 2000),
because, as mentioned above, the geometry of the caustic
is already known up to a scale factor $q^{1/2}$.  The ratio of the
angular size of the caustic to the angular radius of the source
can then be determined from the duration of the caustic crossing
(e.g., Afonso et al.\ 2000; Albrow et al.\ 1999, 2000, 2001).  If
the dereddened color and magnitude of the source is measured (e.g.,
from $JHK$ photometry), then the angular size of the source can
be determined from the color/surface-brightness relation (e.g.,
van Belle 1999). Even with just $HK$ photometry, a good estimate
can be made because, for fixed apparent color and magnitude, the
inferred source size depends only weakly on the reddening. It may
not be possible to obtain $H$ photometry for these very faint stars;
however, the sources at the faint magnitudes at which we expect to
find lenses should be main-sequence stars, and since their distance
will be known, it should be possible to estimate their angular size
even with only the $K$ magnitude.
Finally, since $\gamma$ is known, the ratio of the angular size of the 
caustic to $\theta_\e$ immediately gives the mass ratio, $q$.

        Astrometry provides an independent method to determine $q$
for both caustic and non-caustic crossing events.  For fixed
source trajectory through this geometry, the astrometric deviation is
proportional to $q\theta_\e$. Hence, if this trajectory were known, then 
measurement of the astrometric deviation would yield $q$.

        As discussed above, if the source passes through the caustic, then 
the trajectory can be easily determined from the photometric light curve alone.
However,
if the source passes outside a cusp, the photometric light curve is
potentially subject to two interpretations: the approached cusp could be
along the Sgr-A*/black-hole axis, or perpendicular to it.  These two cases
are nevertheless easily resolved astrometrically since the two 
astrometric deviations are at right angles to one another.  Once it is 
resolved, one knows from the height of the photometric peak of the
deviation, how close the source was to the cusp when it crossed the cusp
axis.  The astrometric deviation at this point then gives $q$.

        For caustic-crossing events the astrometric deviation is of order 
$q\theta_\e\sim 1.5\,{\rm mas}(\theta_\e/1'')$.  Recall that just to
resolve the imaged stars requires {\it resolution} $\ll 0.\hskip-2pt ''1$, so
{\it centroiding} the images to sub-mas levels should be challenging but
perhaps feasible.  However, in the limit of $\gamma\ll 1$ (which dominates
the expected distribution of lenses shown in Fig.\ 2), the maximum
astrometric deviation is $q\theta_\e/\sqrt{8}$.  This factor $\sim 3$
reduction relative to the caustic-crossing case could render an
already difficult measurement impossible.

\section{Discussion}

We have shown that a pair of lensed images from a background bulge star
is likely to be found by the time stars are imaged down to $K=21$ within
an arc second of Sgr A*, and about 8 pairs are expected if $K=23$ is
reached. These images will need to be distinguished from
the much more numerous stars in the vicinity of Sgr A* by finding a
pair of images with the correct relations between magnifications,
positions and proper motions expected for a point mass lens.

The detection of these lenses would show that Sgr A* is concentrated
within a radius $\la 1''$, but this has already been demonstrated at
much smaller radii by the direct measurement of stellar accelerations
(Ghez et al.\ 2000).  A much more interesting application would be to
test the predicted presence of a cluster of stellar black holes around
Sgr A*. Unfortunately, we find that these events should be quite rare:
only about one caustic-crossing event per century for every lensed
image that is monitored. However, this rate would be enhanced if a pair
of highly magnified images were found.  


If an event is detected toward the black-hole cluster, it need not
necessarily be due to a black hole: there will also be events in this
direction caused by ordinary stars around the black-hole cluster.
Assuming a core radius of $r_c = 0.7\,$pc for the old population of
low-mass stars (caused by scatterings from the cluster black holes),
the surface mass density of these stars is a factor
$(R_0\theta_\e/r_c)^{3/4}/0.18\simeq 0.45(\theta/0.\hskip-2pt ''5)^{3/4}$
smaller than that of the black holes.
The duration of these events due to old stars would be similar
to the ones from the cluster black holes, since the lower masses are
roughly compensated by the lower velocity dispersion.  However, as
discussed in \S\ 4, they could be distinguished by photometric
monitoring of a caustic crossing and/or astrometrically.

Alexander \& Loeb (2001) have also investigated microlensing by
stellar objects near Sgr A*. In their case, they consider background
disk stars as sources (whereas we have considered bulge stars), and
they assume that a population of $1 M_{\odot}$ stars surrounds Sgr A*,
with a density profile $\rho\propto r^{-3/2}$. They discuss the
increased likelihood to observe a star with high magnification when
considering the microlensing by stellar objects near the Einstein
radius of Sgr A*.

We note that more microlensing events by cluster black
holes might be found by monitoring to the same depth all the stars
over the entire area of the black hole cluster, within $\sim 20''$ of
Sgr A*. In this case, we would not know a priori which of the source
stars are at a large distance behind the GC, and the magnification by
Sgr A* would be negligible. The events from the cluster black holes
would tend to have Einstein timescales $t_\e$ that were $(m_{\rm
bh}/\overline{m})^{1/2}\sim 6$ times longer than those of ordinary
bulge events (since, far away from Sgr A*, the velocities of the black
holes are similar to those of other bulge stars).  They could not be
unambiguously distinguished on the basis of timescales alone because
the full-width at half-maximum of the mass estimate is almost a factor
100 (Gould 2000).  However, since most of this width is due to the
large dispersion in proper motions $\mu$, it could largely be removed
by measuring the astrometric deviation, which yields the Einstein
radius $\theta_\e$, and so $\mu=\theta_\e/t_\e$.  Since the surface
density of black holes is $\Sb \propto \theta^{-3/4}$, and their
velocity dispersion $\sigma\propto \theta^{-1/2}$, the number of
microlensing events they produce within $\theta$ is proportional to
$\theta^{3/4}$, so there should be about 10 times more events from the
entire cluster, within $20''$, than events that are affected by the
magnification of Sgr A*.

Finally, we point out a very interesting additional application of the
lensed pairs that we predict should be found at any detection limit:
they are potentially powerful probes of the extinction immediately
behind the GC, which is the difference between the extinction toward
the source, and the extinction toward GC sources along the same line
of sight as the images.  As we discuss immediately below, the former
can be determined by obtaining multi-color photometry of both images
and spectroscopy of the brighter one.  The latter can be found from
the distribution of colors and magnitudes of sources along lines of
sight neighboring the images, the great majority of which lie within
$\la$ 1 pc of the GC.  Recall from
\S\ 4,  that  $\theta_\e$ can  be determined  from the measured  image
positions, which implies that $D_{ls}$, the source distance behind the
GC, is known from equation (\ref{eqn:thetaeofd}).

	To accurately measure the extinction to the source, its
intrinsic color must be determined.  This can best be done by 
spectroscopy of the brighter image.  A large fraction of the
lensed sources have small $\gamma$ (see Fig.\ 2), and so consist
of a luminous source whose brighter image is weakly magnified, and whose
fainter image is heavily demagnified.  Hence the brighter image must have
$K\leq K_\thr + 5\log\gamma$.  Thus, these low-$\gamma$ image pairs, which 
contribute
almost nothing to the probability of detecting cluster black holes,
are the best probes of extinction.  See Figure 1 for the overall distribution
of brighter images.  For cases where the brighter image is sufficiently
faint, spectroscopy will no longer be possible, and it will be necessary
to determine the intrinsic color from multi-color photometry.  This
will be less accurate, but probably still useful.  Of course, once the
intrinsic color of the source is known from measurements of the brighter
image, extinction can be measured along {\it both} lines of sight to
the source, by measuring the apparent colors of both images.

\begin{acknowledgements}

JC thanks Scott Gaudi for useful discussions and insights.  Work by AG
was supported in part by grant AST 97-27520 from the NSF, and in part
by a grant from Le Minist\`ere de L'\'Education Nationale de la
Recherche et de la Technologie.

\end{acknowledgements}

\clearpage

\begin{figure}
\vspace*{-1cm}
\plotone{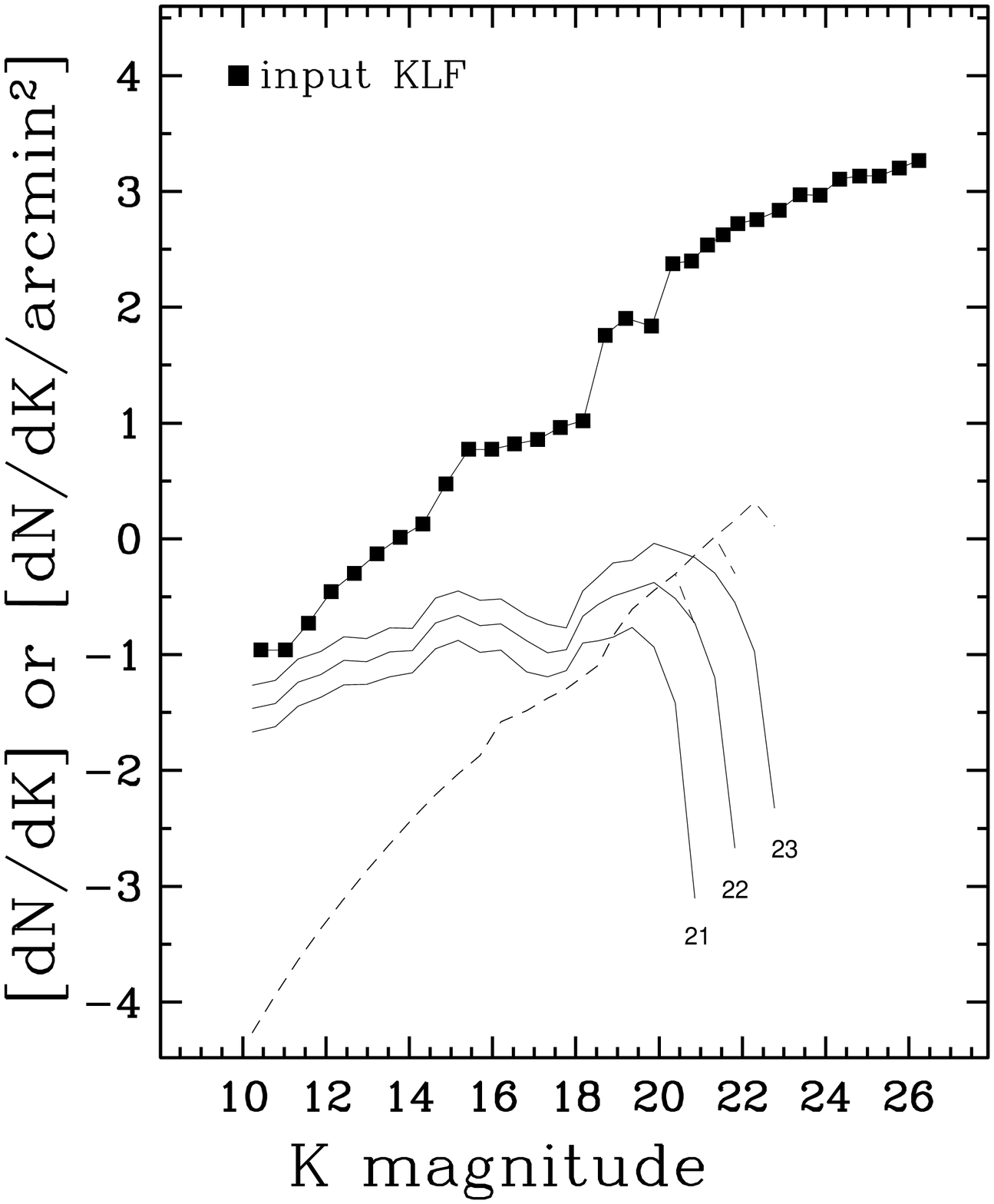}
\vspace*{-3cm}
\caption[junk]{\label{fig:one}
The extincted surface density \,$K=K_{0}+A_K$\, luminosity function
[dN/dK/arcmin$^{2}$] derived from the Zoccali et al.\ (2000) data is
shown as filled squares.  The estimated GC extinction $A_K=3$ has been
added to facilitate direct comparison with the other curves, which all
assume this extinction.  Solid lines represent the number of brighter
images per magnitude [dN/dK] of bulge stars lensed by Sgr A$^{\ast}$
for three detection thresholds.  Similarly, the dashed lines represent
the corresponding distributions of the fainter images.
}
\end{figure}

\clearpage

\begin{figure}
\vspace*{-2cm}
\plotone{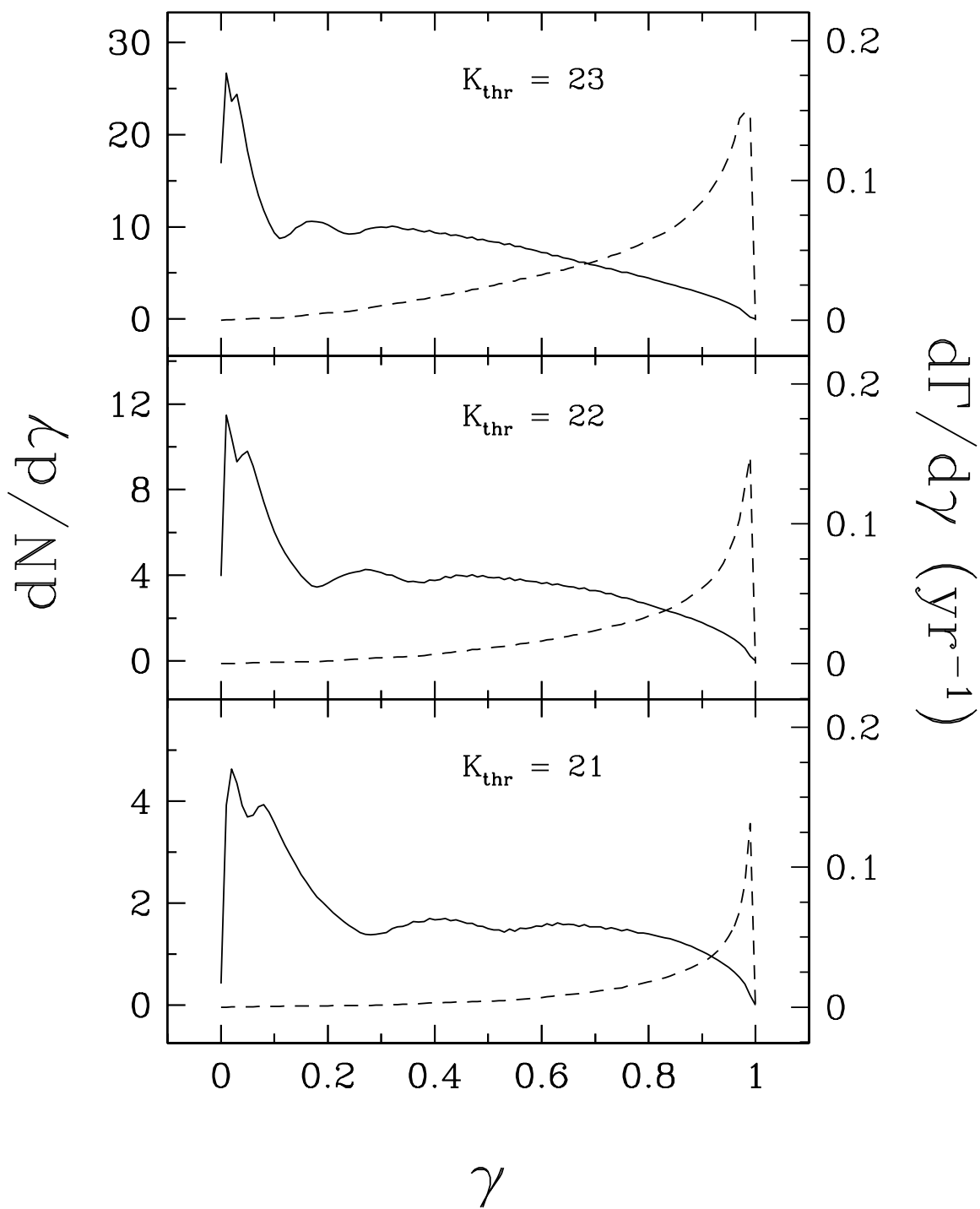}
\vspace*{-1cm}
\caption[junk]{\label{fig:two}
The distributions of the detectable number of sources lensed by Sgr A*
as a function of the shear $\gamma$ (solid lines), and of the rate of
microlensing events produced by the cluster black holes,
d$\Gamma$/d$\gamma$ (dashed lines), for the three magnitude thresholds
mentioned in the text.  Note that d$\Gamma$/d$\gamma$ =
$\omega(\gamma)$ $\cdot$ dN/d$\gamma$, where $\omega(\gamma)$ is given by
eq. (11).
}
\end{figure}

\begin{figure}
\vspace*{-3cm}
\hspace*{-19cm}
\plotfiddle{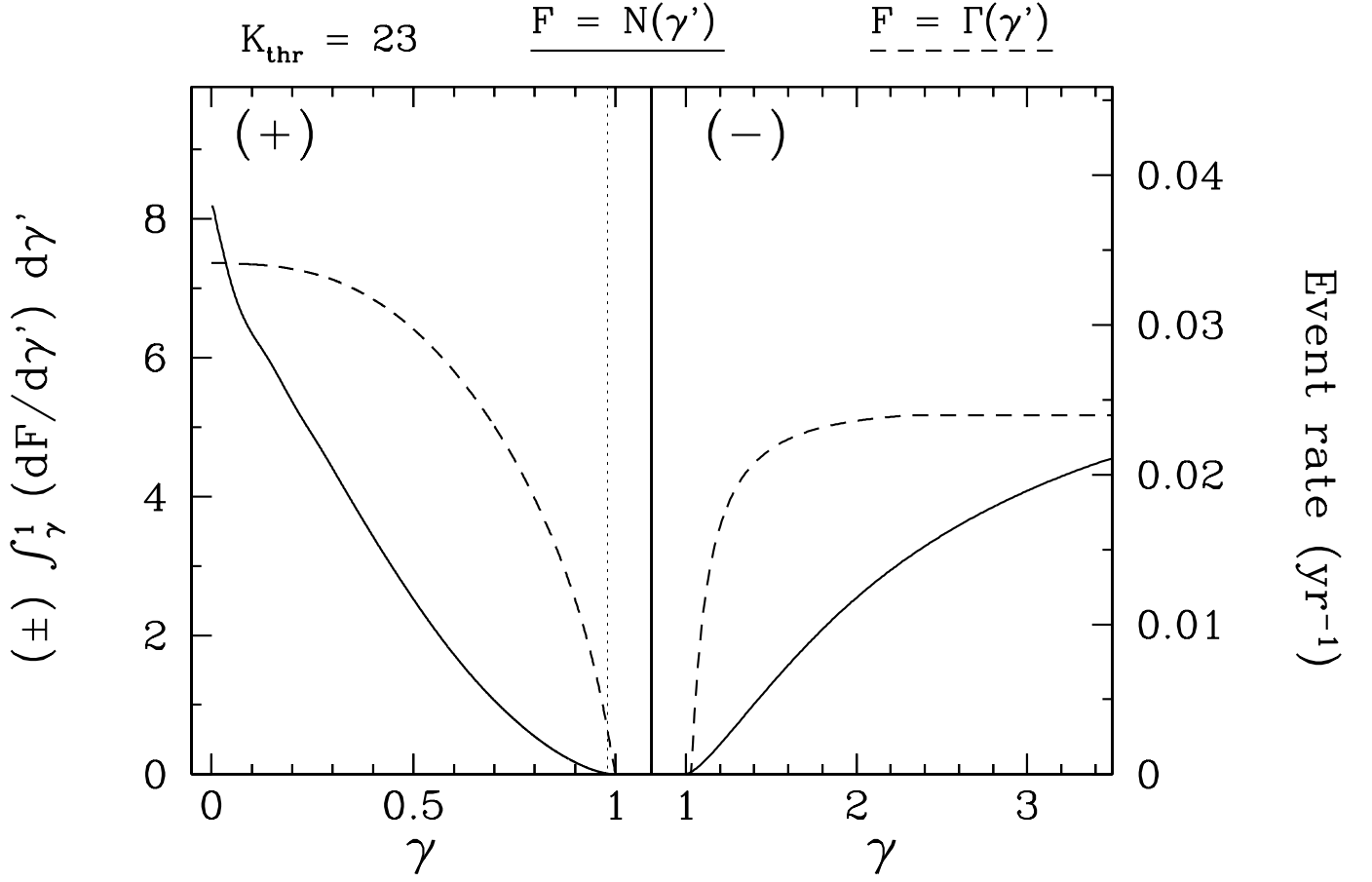}{1cm}{-90}{0.8}{0.8}{0}{0}
\vspace*{-1cm}
\caption[junk]{\label{fig:three}
Cumulative shear distributions of the number of sources lensed by Sgr
A* (solid lines) and of the rate of microlensing events (dashed
lines), for a magnitude limit of $K_{thr} = 23$.  The left panel
corresponds to the brighter pair members, and the right panel to the
fainter ones.  The (plus or minus) sign between parenthesis in each
panel indicates the corresponding sign of the integral in the
left-hand vertical label.  The vertical dotted line in the left panel
marks the point where the Chang-Refsdal approximation begins to break
down (see text).
}
\end{figure}

\begin{figure}
\vspace*{-2cm}
\plotone{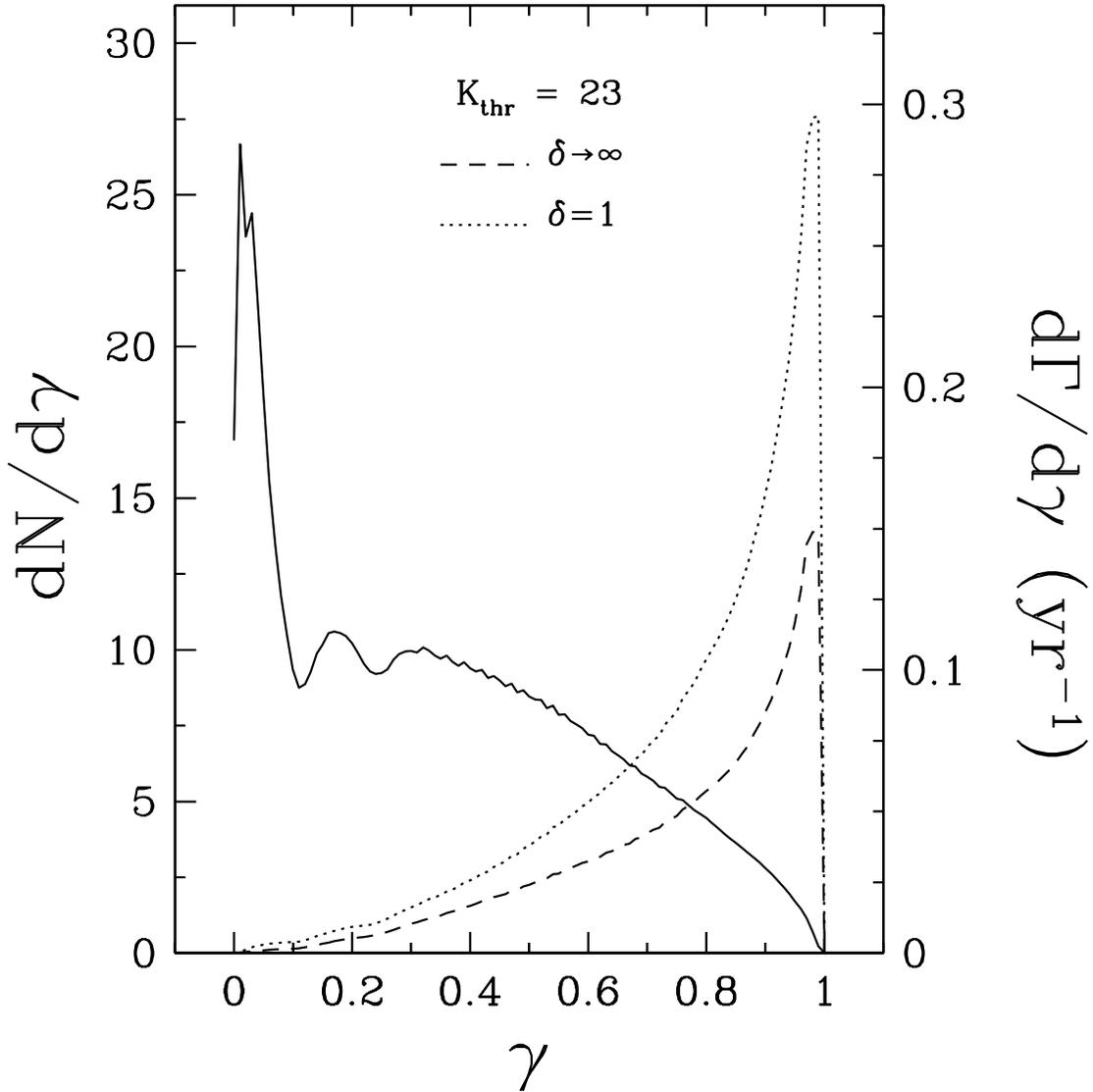}
\vspace*{-1cm}
\caption[junk]{\label{fig:four}
Comparison of the shear distributions of the rate of microlensing
events for the cases when caustic crossing is required (dashed line),
and when a peak magnification of at least a factor of two relative to
the magnification by Sgr A* is required (dotted line).  In the latter
case the total rate (area under the distribution) is almost doubled
compared to the caustic crossing case.  The solid line is the same as
in the upper panel in Figure 2.
}
\end{figure}

\end{document}